# Prompt Sensitivity of Generative Agents: Evidence from an Epidemic Model

Ross Williams, Niyousha Hosseinichimeh

Industrial and Systems Engineering, Virginia Tech; Alexandria, VA 22305, USA.

Email: rossfw@vt.edu

## Abstract

As generative AI gains traction, researchers are investigating its potential to serve as proxies for humans. From undergoing cognitive psychology experiments to experiencing an epidemic, generative agents, agents powered by generative AI models, produce realistic human behavior when prompted. This study explores the sensitivity of these generative agents' behavior to prompt modifications and varied persona names of the agents. To assess this sensitivity, we use a generative agent epidemic model, wherein each agent is prompted daily on whether it wants to isolate or commingle with other agents. We found that using synonymous prompts results in negligible changes to the model's outcomes. However, minor variations in prompts, as well as contextual changes, do influence the model's results. Lastly, our data indicates that different persona names assigned to generative agents, specifically those imbued with personas, do not significantly impact epidemic outcomes.

**One-Sentence Summary:** In an epidemic model, generative agents exhibit behavior that is insensitive to synonymous prompt and persona name modifications but have behavior that is sensitive to slight and contextual prompt modifications.

## 1. Introduction

Over the past year, the study of generative artificial intelligence (AI) has seen significant advances (26). Generative AI refers to AI models that can create content such as text, images, videos, and music in response to user prompts. Since these models are trained on massive amounts of human-generated data (5), social scientists have explored their ability to roleplay as a representative proxy for human behavior.

From exploring tacit collusion (1) to creating more humanistic thinking chatbots (2-8), the possibilities of modeling human behavior using agents powered by large language models (LLM) seems increasingly likely (9). A generative agent is an entity that uses LLMs to personify itself and simulate realistic human behavior (2). To interact with the generative agent, the author feeds text prompts as inputs to the LLM (1, 2, 10–12). The LLM then proceeds to generate text responses as outputs (5). In some applications of generative AI, the survey-like experiment ends there, and the output responses are analyzed (13, 14, 18–20). For other experiments, researchers have conducted dynamic experiments with generative agents on social science models such as information diffusion (2) and epidemic models (10) to test their believability as proxies for human behavior.

For the information diffusion model, generative agents are placed in a virtual, pixelated world to go about their daily lives (2, 26). In this seminal paper, generative agents have names and traits; together, the agents make up a small-town named Smallville. In this community, an agent can interact with other agents, the environment, and even itself. The agent recalls its actions and conversations through a memory architecture, which in turn helps the agent plan and act in future time steps (15). To test information diffusion in this simulation, the authors implanted a thought in a single agent of throwing a Valentine's Day party. With time, the agent—as well as other invited guests—spread the word of the Valentine's Day party, resulting in many agents attending the event.

While the information diffusion model is a proof-of-concept that generative agents can operate in social-science experiments, the epidemic model quantitatively studies an epidemic in Dewberry Hollow, a virtual town full of generative agents (10). Here, agents are informed daily about their own health as well as the fictional virus spreading within their community. Feeding this information in the form of a prompt to ChatGPT, these agents, personified with the Big Five traits, make decisions on whether they want to isolate or commingle. These individuals' decisions exhibited heterogeneous risk-tolerances, and collectively, their behavior creates multiple waves of an epidemic.

Other dynamic models incorporating generative agents have also begun to emerge. In some game theory applications, generative agents compete against each other in the Prisoner's Dilemma (11, 12) as well as the Battle of the Sexes (11). For the study of social networks, researchers have agents dynamically react to each other's posts and comments in a Reddit-like setting (16). Like the Reddit paper, another model named $S^3$ creates a social-network as well as realistic human-personas from data-sources (17). Furthermore, they calibrate the agent-to-agent social media interactions to replicate information diffusion found on social media for two events: the Japan Nuclear Waste Water Release Event and the Chained Eight-child Mother event.

Regardless of whether the generative agent application is static or dynamic, all these models lack one thing: robust sensitivity analysis. As LLMs have grown in popularity, concerns about their sensitivity to text-input have emerged (21, 22). However, according to Wang et al.'s comprehensive literature review of generative agents, there is no unified way to prompt LLMs nor to conduct prompt sensitivity analysis (15). Compared to traditional sensitivity analysis, generative agent sensitivity analysis must have prompt sensitivity analysis in addition to regular parameter sensitivity analysis. Prompt sensitivity analysis has a significantly larger dimensional space due to the infinite possibilities of linguistic expression that can convey the same meaning. Given that LLMs are designed to predict the next word in a sequence, it is critical that variations in prompts do not drastically change the generative agent's behavior. Concerns about sensitivity of prompt modifications stem from researchers post-processing LLM text outputs to determine the decision the generative agent shall take (2, 10, 16, 17); hence, if a minor prompt change impacts the agent decision, the results from the model would not be robust. Therefore, it is vital to assess the sensitivity of these LLMs to different input prompts to understand how it influences the generative agents' behavioral response.

This paper is a step toward generative agent sensitivity analysis. As it is impossible to explore the entire prompt sensitivity space for an LLM, we will use the generative agent epidemic model as an example model to test prompt sensitivity analysis (10). This model was selected amongst other generative agent models for two distinct reasons. First, this model has a mostly non-changing vignette-style prompt fed into a LLM each time step for each agent. Comparatively, some models, like the Valentine's Day information diffusion model, have generative agents with memory architecture, which in turn creates more input prompt variability (1, 2, 16, 17). Therefore, the epidemic model prompt helps reduce the potential amount of prompt dimensionality juxtaposed with other models. Secondly, the epidemic model has an easy to measure impact from the prompt: the resulting mobility curve. Mobility curves (seen in Figure 2 and Figure 3) compare the mobility of people at the current time step to the percent of new active case infections that happened the previous time step. If the prompt is sensitive to alterations, then the resulting mobility curve will have a different shape as well as values. To comprehensively analyze the robustness of the prompt (10), three types of prompt sensitivity analysis will be conducted: semantic variation sensitivity, contextual shift sensitivity, and persona name sensitivity.

Semantic variation sensitivity analysis modifies parts of the prompt by substituting them with terms or phrases that are similar semantically to assess any resulting changes in the LLM's response. When using synonymous phrases in prompts (e.g., "knows about" and "is aware of"), we hypothesize that the agent's behavior will not change; thereby, not altering the mobility curve. In contrast, we hypothesize that using nonsynonymous phrasing (e.g., "knows" and "learns") will result in different agent behavior.

Contextual shift sensitivity analysis involves altering the scenario the generative agent faces by modifying the prompt. To test the epidemic model's contextual shift sensitivity, we examine if agents, facing less fiscal pressure, react differently compared to the baseline prompt. We hypothesize that the resulting agent's behavior will change as the agents confront a different scenario.

Lastly, for persona name sensitivity analysis, generative agents' names are substituted for other names to evaluate the differences in the LLM's response based on name change (e.g. "Betty" versus "Jamal"). A major concern that researchers have when leveraging AI models is perpetuating racial and sex biases (23). In 2015, Amazon shut down its recruiting AI tool as it was found to be biased against women (24). Similarly, in 2020, Google apologized for its Vision AI model mislabeling objects as guns held by darker-skinned individuals, while identifying them as different objects for lighter-skinned individuals (25). Hence, it is vital to ensure LLMs are not biased by generative agents' names: names which may subtly hint at sex and race. We hypothesize that when generative agents have personality traits, such as the Big Five personality traits, their names are not a significant factor. The logic is that human-traits will be the primary driver of how the generative agent acts versus that of their name.

Through several experiments, we present compelling evidence and statistically significant results. For semantic variation sensitivity analysis, we find that synonymous phrasing produces no statistically significant deviations from the baseline. For nonsynonymous phrasing, statistically significant deviations were noted. Similarly, our contextual shift sensitivity analysis showed a statistically significant different mobility curve than the baseline. Lastly, all tests conducted for name sensitivity analysis resulted in no statistically significant deviation from the baseline results.

## 2. Materials and Methods

<u>Background model information:</u>

Since the analysis of this paper relies on Williams et al.'s model, this section provides necessary background information of their generative agent-based epidemic model as well as a summary of some of their findings (10). At the start of each time step in the epidemic model, each generative agent is asked whether it wants to stay at home. To arrive at this decision, each agent is provided a prompt (Figure 1, Panel A) with static information, such as their name, age, personality traits, and job status. The prompt also contains dynamic information about the agent's health and the percentage of new active cases detected in the community during the previous time step. After all the information is supplied, ChatGPT functions as the agent's decision-making ability, responding with either "yes" or "no" to the question of whether the agent wants to stay at home. If the agent opts to stay at home, it will not have any encounters with other agents during that time step. Conversely, if the agent decides to venture outside, it will interact with five other agents. Contacts between infected and susceptible agents result in a probabilistic chance that the susceptible agent becomes infected.

Using this algorithm, the epidemic model produces intriguing outcomes at both individual and societal levels. In simulations that incorporated self-health feedback, many agents displaying symptoms such as fever and cough chose to quarantine, thereby slowing the spread of the epidemic. When the model included both self and societal health feedback, agents not only self-quarantined but also chose to isolate when the community reported lots of new active cases. This behavior contributed to flattening the epidemic curve through self-isolation. The agents, with

**Base prompt** A

You are [agent's name]. You are [agent's age] years old.
Your traits are given below:
[agent's traits]
Your basic bio is below:
[agent's name] lives in the town of Dewberry Hollow. [agent's name] likes the town and has friends who also live there. [agent's name] has a job and goes to the office for work everyday.
I will provide [agent's name] 's relevant memories here:
[agent's health feedback]
[agent's name] knows about the Catasat virus spreading across the country. It is an infectious disease that spreads from human to human contact via an airborne virus. The deadliness of the virus is unknown. Scientists are warning about a potential epidemic.
[agent's name] checks the newspaper and finds 4.4% of Dewberry Hollow's population caught new infections of the Catasat virus yesterday.
[agent's name] goes to work to earn money to support [agent's name] 's self.
Based on the provided memories, should [agent's name] stay at home for the entire day? Please provide your reasoning.

**Is aware of prompt** B

You are [agent's name]. You are [agent's age] years old.
Your traits are given below:
[agent's traits]
Your basic bio is below:
[agent's name] lives in the town of Dewberry Hollow. [agent's name] likes the town and has friends who also live there. [agent's name] has a job and goes to the office for work everyday.
I will provide [agent's name] 's relevant memories here:
[agent's health feedback]
[agent's name] is aware of the Catasat virus spreading across the country. It is an infectious disease that spreads from human to human contact via an airborne virus. The deadliness of the virus is unknown. Scientists are warning about a potential epidemic.
[agent's name] checks the newspaper and finds 4.4% of Dewberry Hollow's population caught new infections of the Catasat virus yesterday.
[agent's name] goes to work to earn money to support [agent's name] 's self.
Based on the provided memories, should [agent's name] stay at home for the entire day? Please provide your reasoning.

**Learns prompt** C

You are [agent's name]. You are [agent's age] years old.
Your traits are given below:
[agent's traits]
Your basic bio is below:
[agent's name] lives in the town of Dewberry Hollow. [agent's name] likes the town and has friends who also live there. [agent's name] has a job and goes to the office for work everyday.
I will provide [agent's name] 's relevant memories here:
[agent's health feedback]
[agent's name] learns about the Catasat virus spreading across the country. It is an infectious disease that spreads from human to human contact via an airborne virus. The deadliness of the virus is unknown. Scientists are warning about a potential epidemic.
[agent's name] checks the newspaper and finds 4.4% of Dewberry Hollow's population caught new infections of the Catasat virus yesterday.
[agent's name] goes to work to earn money to support [agent's name] 's self.
Based on the provided memories, should [agent's name] stay at home for the entire day? Please provide your reasoning.

**Savings prompt** D

You are [agent's name]. You are [agent's age] years old.
Your traits are given below:
[agent's traits]
Your basic bio is below:
[agent's name] lives in the town of Dewberry Hollow. [agent's name] likes the town and has friends who also live there. [agent's name] has a job and goes to the office for work everyday.
I will provide [agent's name] 's relevant memories here:
[agent's health feedback]
[agent's name] knows about the Catasat virus spreading across the country. It is an infectious disease that spreads from human to human contact via an airborne virus. The deadliness of the virus is unknown. Scientists are warning about a potential epidemic.
[agent's name] checks the newspaper and finds 4.4% of Dewberry Hollow's population caught new infections of the Catasat virus yesterday.
Even though [agent's name] already has enough savings, [agent's name] goes to work to earn money.
Based on the provided memories, should [agent's name] stay at home for the entire day? Please provide your reasoning.

Fig. 1. Comparison of prompts used to test semantic variation and contextual shift sensitivity analysis. Panel A is the baseline prompt used. Panel B uses a synonymous prompt by replacing "knows about" with "is aware of". Panel C replaces "knows" with "learns". Panel D changes the context the faces.

various traits, exhibited varied levels of risk-responsiveness, leading to heterogenous behaviors. The collective actions of these agents resulted in multiple waves of the virus and ongoing endemic states.

We use the self and societal health feedback prompt for our prompt sensitivity analysis due to its ability to elicit a diverse set of reasonings from the agents: not only do some ill agents provide reasons to quarantine, but risk-responsive agents also justify their decisions to isolate (10). Since the primary goal of the generative agent epidemic model is to integrate human behavior into epidemic models, it is essential agents are informed about the virus, which only occurs when both self and societal health feedback prompt.

Experimental Design

We ran three types of prompt sensitivity analysis: semantic variation, contextual shift, and persona name sensitivity analysis. Semantic variation sensitivity analysis modifies portions of the prompt by replacing them with semantically similar terms or phrases to assess any resulting changes in the LLM's response. Contextual shift sensitivity analysis also modifies portions of the prompt, but with the goal of revising the scenario the generative agent faces. Lastly, persona name sensitivity analysis substitutes the generative agents' names to investigate whether the LLM has any racial or sex biases. Table 1 shows each experiment tested for each type of sensitivity analysis.

**Table 1 – Sensitivity analysis experiments**

| Sensitivity analysis type | Experiment Number | Baseline Scenario | Experiment scenario |
|---|---|---|---|
| Semantic variation | Experiment #1 | "knows about" | "is aware of" |
| | Experiment #2 | "knows" | "learns" |
| Contextual shift | Experiment #3 | "[agent's name] goes to work to earn money to support [agent's name]" | "even though [agent's name] already has enough savings, [agent's name] goes to work to earn money." |
| Persona name | Experiment #4 | Top 50 male and top 50 female names | Top 100 female names |
| | Experiment #5 | Top 50 male and top 50 female names | Top male and female name |

For semantic variation sensitivity analysis, only the prompt in the epidemic model code was altered (27). To test for semantic sensitivity, Figure 1's panel B and C exhibit the semantic modifications that we did to the epidemic model's base prompt, seen in panel A. To test for synonymous semantic sensitivity, experiment #1 replaces the words "knows about" (Figure 1, Panel A) with "is aware of" (Figure 1, Panel B). For nonsynonymous semantic variation, experiment #2 removes the word "knows" and replaces it with the word "learns" (Figure 1, Panel C). Although the two words seem similar, "learns" implies a recently completed journey of understanding while "knows" refers to having already acquired the knowledge. Therefore, we hypothesize that agents with "learns" will be more risk-averse given their novel familiarity with the virus, thereby decreasing the mobility curve values; this means there will be less mobility for an equivalent percentage of new active case infections when compared to the baseline.

Like the semantic variation experiments, contextual shift sensitivity analysis only modifies the prompt in the code. In Figure 1, experiment #3 has the line "[agent's name] goes to work to earn money to support [agent's name]" (Figure 1, Panel A) substituted for "even though [agent's name] already has enough savings, [agent's name] goes to work to earn money (Figure 1, Panel D)." The goal of this statement is to alter the environmental circumstance the agent faces; knowing it has the financial means to stay at home, the agent may be less inclined to go to work. Therefore, we hypothesize the agents will be much more risk-averse as they do not have the fiscal incentive to go to work.

In persona name sensitivity analysis, the prompt remains untouched, but the method to generative names has changed. Experiment #4 assesses sex biases in the LLM. Instead of employing the 50 most popular male and female American names, it uses the 100 most popular female American names. Experiment #5 evaluates the impact of specific individual names. For this, all 100 agents are assigned top male and female names: 50 agents are named Jose, while the remaining 50 are named Maria.

Apart for the experiment changes listed in Table 1, the code closely adheres to the original generative agent epidemic model, though with some adjustment. Consistent with the Williams et al. paper, every run in this study involves 100 agents: 98 susceptible and 2 initially infected. Other parameters are set as follows: the time to heal is 6 days, the contact rate is 5, infectivity is 0.1, and the max number of time steps is 50. However, diverging from the Williams et al. methodology, for the semantic variation and contextual shift sensitivity analyses, we will select the agents' names from the top 100 most popular American names (50 male and 50 female) instead of the top 200 most popular American names (100 male and 100 female). For our study, each experiment will be run 5 times.

## 3. Results

Figure 2 and Figure 3 display the resulting mobility curves from each experiment. To discern differences in these mobility curves relative to the baseline, we employ multilinear regression. The outcomes are presented in Table 2. Here, mobility is the dependent variable as a function of new active daily cases, new active daily cases squared, and a variable called experiment. Experiment is a dummy variable; when it is set to 1 if the data is from the experiment being tested and equals 0 if the data is from the baseline. A statistically significant $\beta_3$ or $\beta_4$ suggests that the experiment has a measurable effect on the resulting mobility curve relative to the baseline.

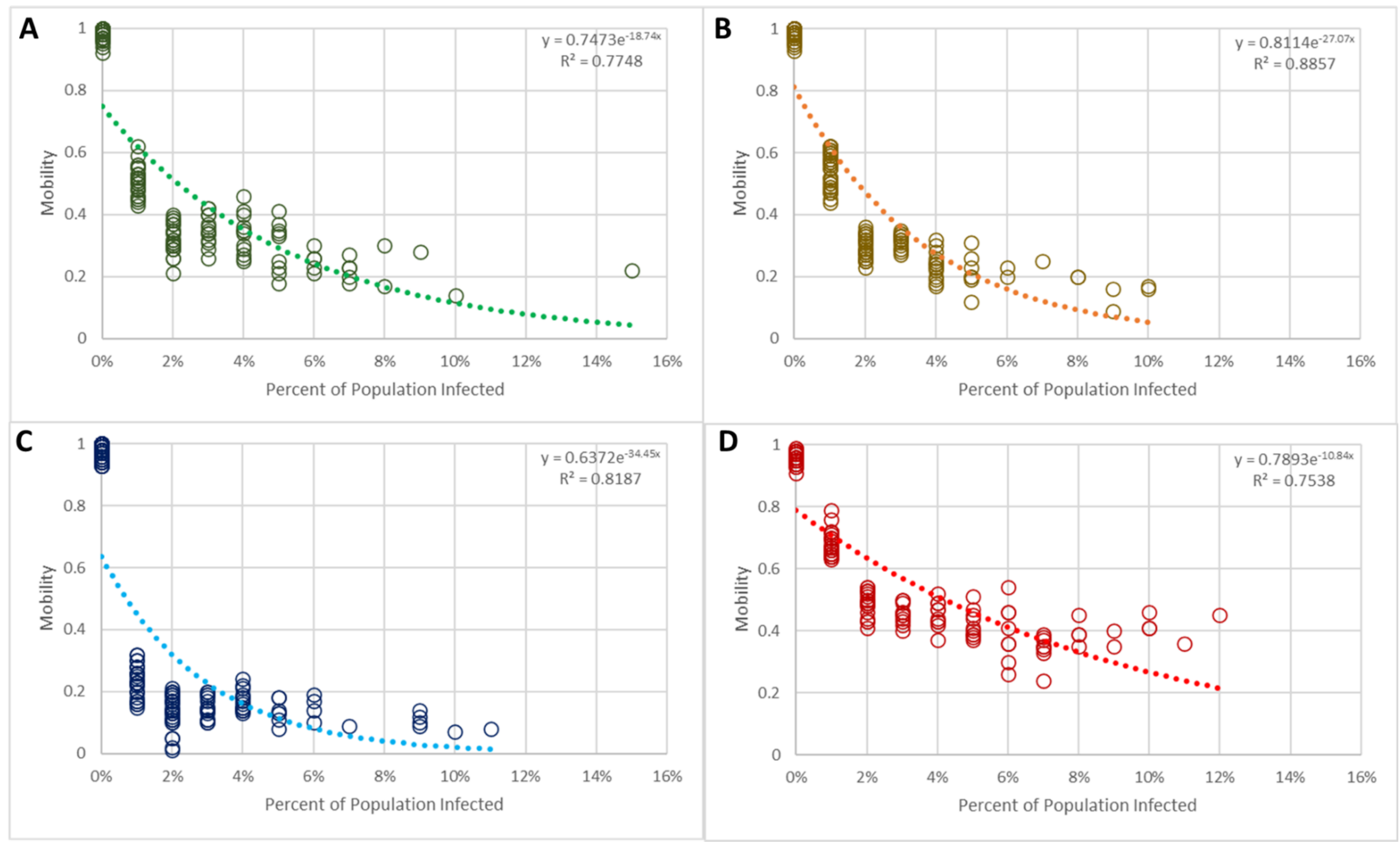

Fig. 2. The mobility curves above are from the semantic variation and contextual shift sensitivity analysis results. Panel A is the baseline mobility curve. Panel B is experiment #1's mobility curve ("is aware of"). Panel C is experiment #2's mobility curve ("learns"). Lastly, Panel D is experiment #3's mobility curve ("even though [agent's name] already has enough savings, [agent's name] goes to work to earn money").

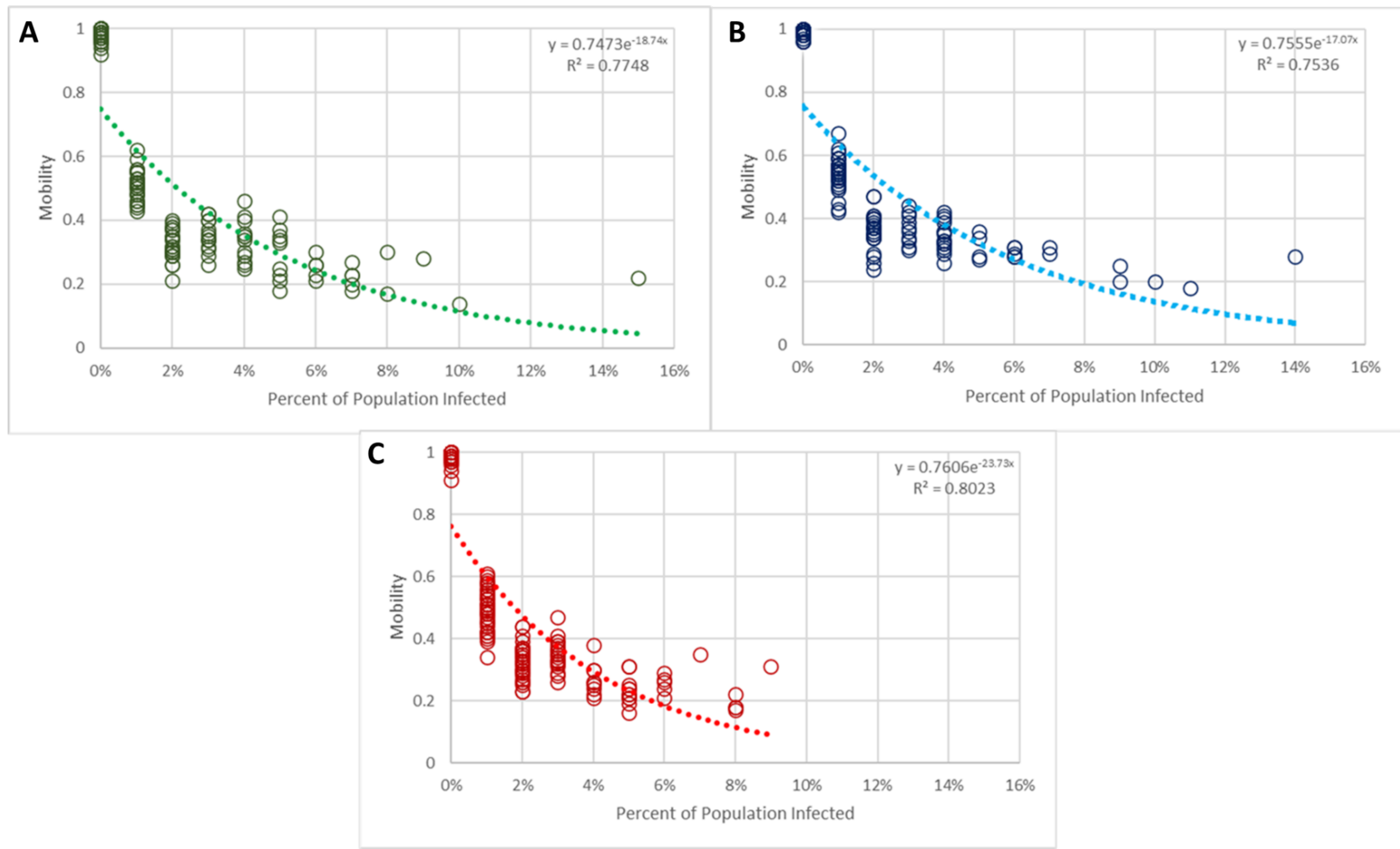

Fig. 3. The mobility curves above are from the persona name sensitivity analysis results. Panel A is the baseline mobility curve. Panel B is experiment #4's mobility curve (top 100 female names). Panel C is experiment #5's mobility curve (top male and female name).

From the results in Table 2, the multilinear regression shows that experiment #1, #4 and #5 are not significant. Therefore, synonymous semantic variations as well as the persona name sensitivity modifications did not impact the model's outcome. For nonsynonymous semantic variation, experiment #2 illustrates differences in semantics matter. Moreover, our hypothesis is affirmed: agents who "learns" about the epidemic are more risk-adverse than the baseline (Figure 1, Panel A). For contextual shift sensitivity analysis, experiment #3 reaffirms that changing the scenario agents face does result in a different mobility curve. However, unlike our hypothesis, agents who have "enough savings" display less risk aversion. Rather than living off their savings, these agents are more inclined to work compared to those who must support themselves.

## 4. Discussion

In this paper, we offer a first step towards prompt sensitivity analysis for generative agent-based modeling. We explored three different types of sensitivity analysis: semantic variation, contextual shift, and persona name sensitivity analysis. For the semantic variation sensitivity analysis, we demonstrated that exchanging synonymous phrases in a prompt, "knows about" and "is aware of", led to no discernible change in the mobility curve. However, we observed that nonsynonymous semantic variations do impact the results of the model: the "learns" run exhibits reduced mobility compared to the "knows" run. When we made changes to the prompt that shifted the scenario faced for the agent, we noted departures from baseline behavior.

Contrary to our initial hypothesis, the contextual shift sensitivity analysis resulted in more mobile agents. It is unclear why ChatGPT deduced that individuals who opt to work, despite having enough savings, remain more mobile during an epidemic. One hypothesis is that the LLM assumes that agents who already go to work at their own volition, without fiscal imperative, do in fact enjoy going to work.

Our persona name sensitivity analysis indicated that agents' names did not statistically influence the model's results. This finding supports our earlier assertion that persona names, when coupled with personality traits, do not have a major influence on the epidemic outcomes simulated by our model.

Given these findings, this research contributes to the literature on generative agent-based modeling, introducing a first step towards prompt sensitivity analysis (15). While traditional agent-based models have long emphasized the importance of sensitivity analysis (29), the advent of generative agent-based modeling necessitates new methodologies to address the unique challenges of exploring the dimensional space of prompt sensitivity. The findings of this paper have the potential to influence how other modelers approach prompting generative agents and prompt sensitivity analysis in their generative agent-based models.

Beyond impacting the field of generative agent-based modeling, this study also contributes to the field of epidemiology through investigating the robustness of the generative agent epidemic model. Before our work, this model had no sensitivity analysis regarding its prompt. Now, we know prompt changes in Williams et al.'s model affect the results. However, the general behavior responding to cases remains robust. Hence, epidemiological modelers may have more confidence in the base model.

**Table 2 – Experiment multilinear regression results**

| Experiment Number | Experiment #1 Regression | | Experiment #2 Regression | | Experiment #3 Regression | | Experiment #4 Regression | | Experiment #5 Regression | |
|---|---|---|---|---|---|---|---|---|---|---|
| Constant | 0.91*** | 0.91*** | 0.94*** | 0.92*** | 0.85*** | 0.88*** | 0.88*** | 0.88*** | 0.89*** | 0.90*** |
| New Cases | -24.54*** | -23.98*** | -27.03*** | -25.8*** | -19.02*** | -20.68*** | -21.27*** | -21.32*** | -23.05*** | -23.76*** |
| New Cases^2 | 172.82*** | 170.69*** | 196.66*** | 192.91*** | 131.69*** | 131.24*** | 138.87*** | 138.84*** | 164.49*** | 168.07*** |
| Experiment | -0.015 | -0.00 | -0.10*** | -0.07** | 0.09*** | 0.02 | 0.00 | 0.00 | -0.02 | -0.04 |
| New Cases * Experiment | | -0.84 | | -1.80* | | 3.04*** | | 0.10 | | 1.03 |
| R2 | 0.83 | 0.83 | 0.75 | 0.76 | 0.8 | 0.82 | 0.79 | 0.82 | 0.79 | 0.79 |
| N | 383 | 383 | 381 | 381 | 316 | 316 | 328 | 328 | 355 | 355 |

***P<0.000, **P<0.01. *P<0.05

This study has several limitations. First, a significant limitation of this study is that it is impossible to have a prompt sensitivity analysis that encompasses all potential modifications to the prompt. For example, our synonymous test used a single, very close pair ("knows about" versus "is aware of"); more distant synonyms could still shift agent behavior. Although there have been attempts to robustly test generative agent prompts (28), more prompt sensitivity analysis can always be done. To fill this gap, future research can investigate a more systematic, comprehensive list of different types of prompt sensitivity analysis that should be performed. This way, researchers can execute specific types of prompt sensitivity tests instead of heuristically examining the feasible prompt space. Exacerbating this limitation, is the fact that generative agent-based models are resource-intensive. As of June 2023, executing a model of 100 agents across 50 time steps can take over 2 hours of program runtime on a 32GB RAM CPU and incurs a cost of approximately $2 per run. This cost stems from the thousands of application program interface calls made to OpenAI's servers. Compared to traditional agent-based models, the scalability of a prompt sensitivity analysis can become prohibitively expensive. However, as LLMs become more efficient and affordable in the future, the fiscal, computational, and temporal resources of generative agent-based modeling will become less burdensome. Another limitation is that the results may be specific to ChatGPT, and possibly only to that particular version of ChatGPT. Future studies can investigate whether these findings apply beyond ChatGPT to other LLMs.

Overall, this study, by performing prompt sensitivity analysis on a generative agent epidemic model, has demonstrated the robustness of generative AIs' capability to guide dynamic models of human systems. This paper's findings may be used as practical guidance for those working with generative agent-based models, particularly when crafting prompts for generative agents. Looking forward, we hope our results serve as a guiding light for future researchers, aiding them in prompting and in formulating more robust ways to conduct prompt sensitivity analysis for generative agent-based modeling.

## Acknowledgments

**Funding:** This research is funded by US National Science Foundation, Division of Mathematical Sciences & Division of Social and Economic Sciences, Award 2229819.

**Competing interests:** Authors declare that they have no competing interests.

**Data and materials availability:** All data, data-processing and analysis code, as well as the full model, its associated files, and results files, are available online at https://github.com/bear96/GABM-Epidemic and https://github.com/RossFW/Paper2-Prompt-Sensitivity-of-Generative-Agents.